\documentclass[a4paper,11pt]{article}

\usepackage{contribution}

\newcommand{\weblink}[2][]{%
    \ifthenelse{\equal{#1}{}}%
    {\textnormal{\url{#2}}}%
    {\textnormal{\href{#2}{#1}}}%
}

\newcommand{\acknowledgements}[1]{%
  \bigskip\bigskip
  \textsf{\textbf{\Large Acknowledgements}} \\[2ex]
  {#1}
  \bigskip
}

\def\beq{\begin{equation}}
\def\eeq#1{\label{#1}\end{equation}}
\def\eeqn{\end{equation}}

\def\beqa{\begin{eqnarray}}
\def\eeqa#1{\label{#1}\end{eqnarray}}
\def\eeqan{\end{eqnarray}}

\let\bar=\overbar

\def\etal{{\it et al.}}

\def\eg{{\it e.g.}}

\def\half{\frac{1}{2}}

\def\Dslash{\not{\hbox{\kern-4pt $D$}}}
\def\dslash{\not{\hbox{\kern-2pt $\del$}}}

\def\alphas{\alpha_s}
\def\msb{{\bar{\ssstyle M \kern -1pt S}}}

\newcommand{\contribution}[7][]{%
  \clearpage
  \thispagestyle{plain}
  \ifthenelse{\equal{#1}{}}
  {\hypersetup{pdftitle={#2}}}
  {\hypersetup{pdftitle={#1}}}
  \hypersetup{pdfauthor={{#3} {#4}}}
  {\centering\normalfont\LARGE\bfseries\sffamily #2 \par\nobreak}
  \lhead{}
  \chead{%
    \textit{\footnotesize XIV International Conference on Hadron Spectroscopy
      (\weblink[\textit{hadron2011}]{http://www.hadron2011.de}), 13-17 June 2011, Munich, Germany}%
  }
  \rhead{}
  \bigskip
  \begin{center}
    {#3} {#4}\ifthenelse{\equal{#6}{}}{}{\footnote{\weblink[#6]{mailto:#6}}}
    \ifthenelse{\equal{#7}{}}{}{#7} \\
    \textit{#5}
  \end{center}
  \bigskip
}

\renewcommand{\abstract}[1]{%
  \begin{center}
    \begin{minipage}{0.85\textwidth}
      \begin{footnotesize}
        #1
      \end{footnotesize}
    \end{minipage}
  \end{center}
  \bigskip
}

\begin{document}

%
%
%
%
%
{  


%

\newcommand{\smgg}{\ensuremath{\mathrm{SU(3)_c} \otimes \mathrm{SU(2)_L} \otimes \mathrm{U(1)}_Y}}
\newcommand{\mev}{\hbox{ MeV}}
\newcommand{\gev}{\hbox{ GeV}}
\newcommand{\tev}{\hbox{ TeV}}
\newcommand{\jpsi}{\ensuremath{J\!/\!\psi}}
\newcommand{\cgg}{\ensuremath{\mathrm{SU(3)_c}}}
\newcommand{\ncgg}{\ensuremath{\mathrm{SU}(N)_{\mathrm{c}}}}
\newcommand{\ewgg}{\ensuremath{\mathrm{SU(2)_L} \otimes \mathrm{U(1)}_Y}}
\newcommand{\wigg}{\ensuremath{\mathrm{SU(2)_L}}}
\newcommand{\ygg}{\ensuremath{\mathrm{U(1)}_Y}}
\newcommand{\emgg}{\ensuremath{\mathrm{U(1)_{EM}}}}
\newcommand{\sLj}[3]{{}^{#1}\!#2_{#3}}
\newcommand{\slj}[3]{\mbox{$^{#1}${\ifcase#2\or S\or 
	 P\or D\or F\or G\fi}$_{#3}$}}

\contribution[The Future of Hadrons]  
{The Future of Hadrons:\\ The Nexus of Subatomic Physics}  
{Chris}{Quigg}  
{Theoretical Physics Department \\
  Fermi National Accelerator Laboratory \\
  P.O.~Box 500, Batavia, Illinois 60510 USA}  
{quigg@fnal.gov \hfill \fbox{FERMILAB-CONF-11-492-T}}  
{}  
%

\abstract{%
  I offer brief observations on matters discussed at the XIV International Conference on Hadron Spectroscopy and explore prospects for hadron physics.
}
%

\section{Impressions and Musings}
The organizers of \textit{Hadron 2011} have asked me to offer some perspectives on the state and prospects of hadron physics, stimulated by the presentations at this meeting and by some themes in the recent literature. I must begin by noting the enormous diversity and reach of experimental programs reported here in Munich, along with quite remarkable progress in theory, including the emergence of lattice QCD as a versatile and reliable tool. The coherence provided by quantum chromodynamics means that insights may arise from unexpected quarters~\cite{adscft}. It is more than ever advisable to take a broad view that integrates across hadronic physics and to connect with the rest of subatomic physics. Connections work both ways: \textit{You} may  answer  questions that seem far afield!
 It also seems important to look  beyond nuclear and particle physics for insights into scattering and bound-state problems,  to seek new ways to address hadronic questions, and to ask, ``How are we prisoners of conventional thinking?'' Part of the pleasant challenge of the moment is that recent advances leave us with many puzzles and opportunities, including ``simple'' questions that we cannot yet answer.

In contrast to biological evolution, lines in theoretical physics that do not win are not  extinguished, never to rise again. We are free to borrow potent ideas from the past and to apply them in new settings, to powerful effect. A recent example is the application of $S$-matrix style unitarity to the evaluation of multiparton amplitudes~\cite{QCDmethods}.  Tools that organized the exploratory particle-production data in the 1970s, such as triple-Regge analysis~\cite{Detar:1971gn}, may offer ways to understand the far more comprehensive measurements now becoming available. If the old ways do not explain every feature of the new data, they should at least show us which results are surprising, and which should be expected.

\section{QCD: the basis of hadronic physics}
Our picture of matter is built on a set
of pointlike ($r \lesssim 10^{18}\hbox{ m}$) constituents: the quark and lepton families. The fundamental fermions interact through forces derived from  \smgg\ gauge symmetries. Quarks are influenced by
the strong interaction, and so carry color, the strong-interaction
charge, whereas  leptons do not feel the
strong interaction, and are colorless.

We have accumulated persuasive evidence---both theoretical and experiments---that quantum chromodynamics underlies all strong-interaction phenomena~\cite{Kronfeld:2010bx}. The defining attribute of QCD is that it is an asymptotically free theory, by which we understand that the coupling strength, or effective charge, diminishes at short distances or high momentum transfers. Determinations of $\alphas$ in a variety of experimental settings confirm the evolution predicted by QCD: the compilation in Figure~\ref{fig:asf} shows that, to excellent approximation, $\alphas(Q)$ varies linearly with $\ln{Q}$, as expected at leading order in perturbation theory. 
\begin{figure}[tb]
  \begin{center}
{\includegraphics[width=0.65\textwidth]{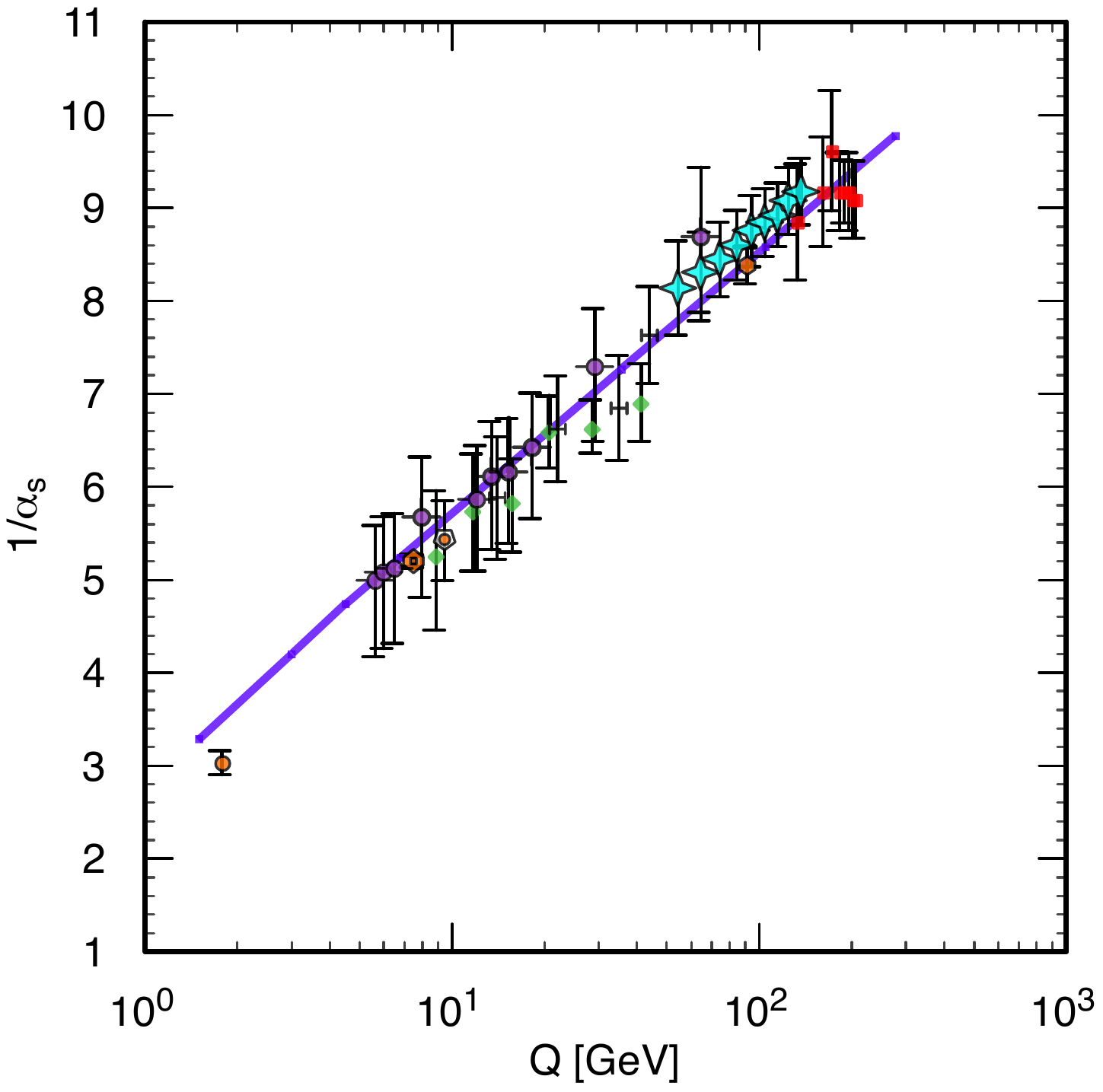}}
    \caption{Measurements of the strong coupling $1/\alphas(Q)$ as a function of the energy scale $\ln{Q}$. The curve shows the QCD prediction at four loops, using 3-loop threshold matching at the heavy-quark pole masses
$m_c = 1.5\gev$, $m_b = 4.7\gev$. From Ref.~\cite{Kronfeld:2010bx}.}
    \label{fig:asf}
  \end{center}
\end{figure}

Over a few decades in the scale $Q$, $\alphas$ decreases from a value at which perturbation theory is a questionable tool to values for which perturbative calculations appear eminently plausible. At still lower scales than those depicted in Figure~\ref{fig:asf}, nonperturbative analysis is required. At high scales and small values of $\alphas$, quarks and gluons, the \textit{fundamental fields} of QCD, are manifest, as we see in high-resolution, hard-scattering studies of proton structure, in phenomena involving matter at high density, and in the formulation of lattice calculations. At lower scales, it is more efficient to deal with \textit{effective} degrees of freedom such as constituent quarks and Goldstone bosons, to formulate effective field theories, and to construct informed approximations such as isobar (resonance) models and semiphenomenological descriptions of nuclei and nuclear structure.

From the variation of $\alphas(Q)$, we see how a theory based on an interaction Lagrangian that contains no dimensionful parameters can be characterized by a scale $\Lambda$, with dimensions of energy, by means of \textit{dimensional transmutation.} The identification
\begin{equation}
{1}/{\alpha_{\mathsf{s}}(2m_c)} \equiv \frac{27}{6\pi}\ln\left(\frac{2m_c}{\Lambda}\right)
\label{eq:LamQCD}
\end{equation}
sets the value of $\Lambda$ appropriate for three active quark flavors. The coefficient of $\ln{Q}$ in the evolution of $1/\alphas$ decreases by $2/6\pi$ at each successive quark threshold.

Lattice QCD including dynamical quarks allows us to evaluate the proton mass according to Einstein's first formulation of his famous equation, which identifies mass with rest energy: $M_p = E_0/c^2$. To excellent approximation, the proton mass is given by a calculable constant times $\Lambda$, plus small corrections~\cite{Durr:2008zz}. Indeed, the quark masses contribute but a few percent of the proton mass: $3(m_u + m_d)/2 \approx 10 \pm 2\mev$~\cite{ishik}.
Hadrons are a thus new kind of matter, whose masses are not given by the sum of the masses of the parts. Instead, the rest energy of a hadron is determined mainly by  confinement energy stored up in the gluon field or the kinetic energy of the quarks~\cite{Quigg:2007dt}.

Since the slope of $1/\alphas(Q)$ changes when new colored objects come into play, we should be alert for slope changes more pronounced than those implied by a new flavor of color-triplet quark. A provocative possibility arises in a simplified view of a supersymmetric world in which an entire family of superpartners becomes active at $1\tev$. The resulting slope change is dramatic: from $7/2\pi$ below the SUSY threshold to $3/2\pi$ above.
It is important to contemplate how the LHC experiments might extend $\alphas$ determinations to much higher scales (perhaps by measuring $W + \hbox{jets}$ ?) and what calculational advances will be needed to enable precise determinations.

\section{Connections and Controlled Approximations \label{sec:control}}
The essence of doing science consists in \textit{making connections} 
that lead us beyond independent explanations for distinct phenoma 
toward a coherent understanding of many phenomena. A network of 
understanding helps us see how different observations fit together 
and---very important---helps us to recognize when 
something \textit{doesn't fit.}

Connections among experiments or observations are not the only
important ones.  Whenever possible, we need to make connections
between our models and the QCD Lagrangian---either directly, or through
effective field theories, lattice field theory, or a controlled
approximation to full QCD. I would also stress the potential value of 
reaching toward connections with our knowledge of nuclear forces, 
with the behavior of nuclear matter under unusual conditions, and with phenomena such as Feshbach resonances~\cite{Feshbach}  encountered in the study of many-body systems~\cite{Braaten:2004rn}.

Under different circumstances,  various approximations to QCD emerge
as controlled expansions in small parameters.
Nonrelativistic QCD applies to heavy-heavy ($Q_{1}\bar{Q}_{2}$) mesons, for which the
quark masses greatly exceed the QCD scale parameter, $m_{Q_{i}} \gg
\Lambda_{\mathrm{QCD}}$~\cite{Brambilla:1999xf}.   NRQCD takes as its expansion parameter the heavy-quark velocity divided by the speed of light.  Heavy-quark effective theory 
(HQET) applies usefully to heavy-light ($Q\bar{q}$) systems, for which $m_{Q}
\gg \Lambda_{\mathrm{QCD}}$~\cite{HQET}.  In first approximation, the spin of the
heavy quark is static, so the ``light-quark spin''
$\vec{\jmath}_{q} = \vec{L}+ \vec{s}_{q}$ is a good quantum number and the
relevant expansion parameter is {$\Lambda_{\mathrm{QCD}}/m_{Q}$}.
Chiral symmetry  is a valuable starting point for light quark systems
($q_{1}\bar{q}_{2}$)~\cite{chidyn}. In 
this case, the expansion parameter compares the current-quark mass to the scale 
of chiral-symmetry breaking, and is generally taken as 
$m_{q}/4\pi f_{\pi}$, where $f_{\pi}$ is the pion 
decay constant. 

\section{What Is a Proton?}
\subsection{Beyond the parton-model idealization \label{subsec:bpmi}}
For hard-scattering phenomena, we have learned to regard the proton as a broadband, unseparated beam of quarks, antiquarks, gluons, and perhaps other constituents, characterized by parton densities $f_{i}^{(a)}(x_{a},Q^{2})$, the number density of species $i$ with momentum fraction $x_a$ seen by a probe with resolving power $Q^2$. The $Q^2$-evolution of the parton densities is given by QCD perturbation theory, but the low-scale distributions $f_{i}^{(a)}(x_{a},Q_0^{2})$ have a nonperturbative origin. Historically, the parton distributions were idealized as free of correlations, and depending only on longitudinal degrees of freedom. Remarkably, this truncated version of reality has served us extraordinarily well for more that forty years.

Intense experimental and theoretical activity on generalized parton distributions~\cite{GPD}, transverse momentum distributions~\cite{Collins}, and the three-dimensional structure of the proton~\cite{Boer:2011fh} represents an important step beyond the traditional idealization. Bjorken~\cite{Bj2010} has suggested that by studying event structure in high-energy collisions we might be able to see signs of correlations among the partons, perhaps by observing diquark--diquark collisions. I would like to underline the importance of looking at both events and distributions for lightly triggered events. At lower energies, the bulk of particle production came from soft collisions---diffraction plus short-range order---as we learned from systematic studies inspired by Ken Wilson~\cite{wilson}. I think it is highly likely that novel structures, and perhaps new classes of processes, will reveal themselves in lightly triggered events at the LHC~\cite{Quigg:2010nn}.

\subsection{Seeking the relevant degrees of freedom: diquarks?}
Much of our insight into how hadrons behave follows from the 
idealization that mesons are quark--antiquark states, 
baryons are three-quark states, and that the quarks have only 
essential correlations. In the case of baryons, this reasoning leads 
us to the plausible starting point of SU(6)  (flavor-spin) wave functions, 
which indeed offer a useful framework for discussing magnetic moments 
and other static properties. Some long-standing observations, however, show us the 
limitations of the zeroth-order guess. If we examine deeply inelastic 
scattering in the limit as $x \to 1$, {spin asymmetries} indicate 
that the SU(6) wave functions are inadequate~\cite{Hughes:1983kf}, and the ratio $F_{2}^{n}/F_{2}^{p}$ is far from the uncorrelated expectation of 
$\frac{2}{3}$~\cite{Melnitchouk:1995fc}.

When we try to account for correlations, it may be fruitful to
consider diquarks as physical objects.  Color
algebra  tells us that the $\mathbf{3} \otimes
\mathbf{3}$ quark--quark combination is attractive in the
$\mathbf{3^{*}}$ representation that corresponds to an antisymmetric
diquark structure.  A simple analysis suggests that the attraction of
$[qq]_{\mathbf{3^{*}}}$ is half as strong as that of the
$[q\bar{q}]_{\mathbf{1}}$ ($\mathbf{3} \otimes \mathbf{3^{*}} \to
\mathbf{1}$) channel.  The suggestion~\cite{Jaffe:1976ig} to
regard members of the scalar nonet  as $qq\bar{q}\bar{q}$ states organized as
$[[qq]_{\mathbf{3^{*}}}[\bar{q}\bar{q}]_{\mathbf{3}}]_{\mathbf{1}}$ is well known.
A few years ago, Selem and Wilczek revisited the
Chew--Frautschi systematics of $N$ and $\Delta$
resonances~\cite{Selem:2006nd}, and found it plausible to
view even low-spin, light baryons as $q[qq]_{\mathbf{3^{*}}}$
configurations.  But now Edwards, \etal~\cite{Edwards:2011jj}, using lattice techniques with dynamical fermions to investigate excited-state baryon spectroscopy,  find $N^*$ and $\Delta$ level structures that do not support a dominant quark--diquark configuration. It is interesting to inquire how 
the quark--diquark picture might square with intuition from the
$1/N_{\mathrm{c}}$ expansion~\cite{Dashen:1993jt}.

It is worth examining the $q[qq]_{\mathbf{3^{*}}}$ proposal
by considering its implications for doubly heavy ($QQq$) baryons.  The
comparison with heavy-light ($Q\bar{q}$) mesons offers a chance to
calibrate the attractive forces in the $\mathbf{3^{*}}$ and
color-singlet channels.  Similarly, extending studies of the
systematics of $qq \cdot \bar{q}\bar{q}$ states to $Qq \cdot
\bar{Q}\bar{q}$ states could develop and
challenge the way we think about diquarks and test color-spin symmetry as an organizing principle.  Finally, in heavy-ion
collisions, we should be alert for tests of the utility of diquarks in
color--flavor locking, color superconductivity, and other novel
phenomena.

\subsection{What are the appropriate degrees of freedom at long distances? \label{subsec:freezing}}
As its momentum decreases, the current quark of perturbative QCD evolves into a constituent quark, as illustrated in Figure~\ref{fig:uni}~\cite{Bhagwat:2007vx}. The constituent-quark mass arises from the cloud of low-momentum gluons winking in and out of existence through quantum fluctuations around the current quark. Dynamical chiral symmetry breaking is an inherently nonperturbative effect that spontaneously generates a quark mass, even in the chiral limit $m \to 0$. This evidence gives a quantitative basis for the highly useful chiral-quark model~\cite{Manohar:1983md}, in which constituent quarks and Goldstone bosons are taken as the apt degrees of freedom at low scales.
\begin{figure}[tb]
  \begin{center}
\includegraphics[width=0.6\textwidth]{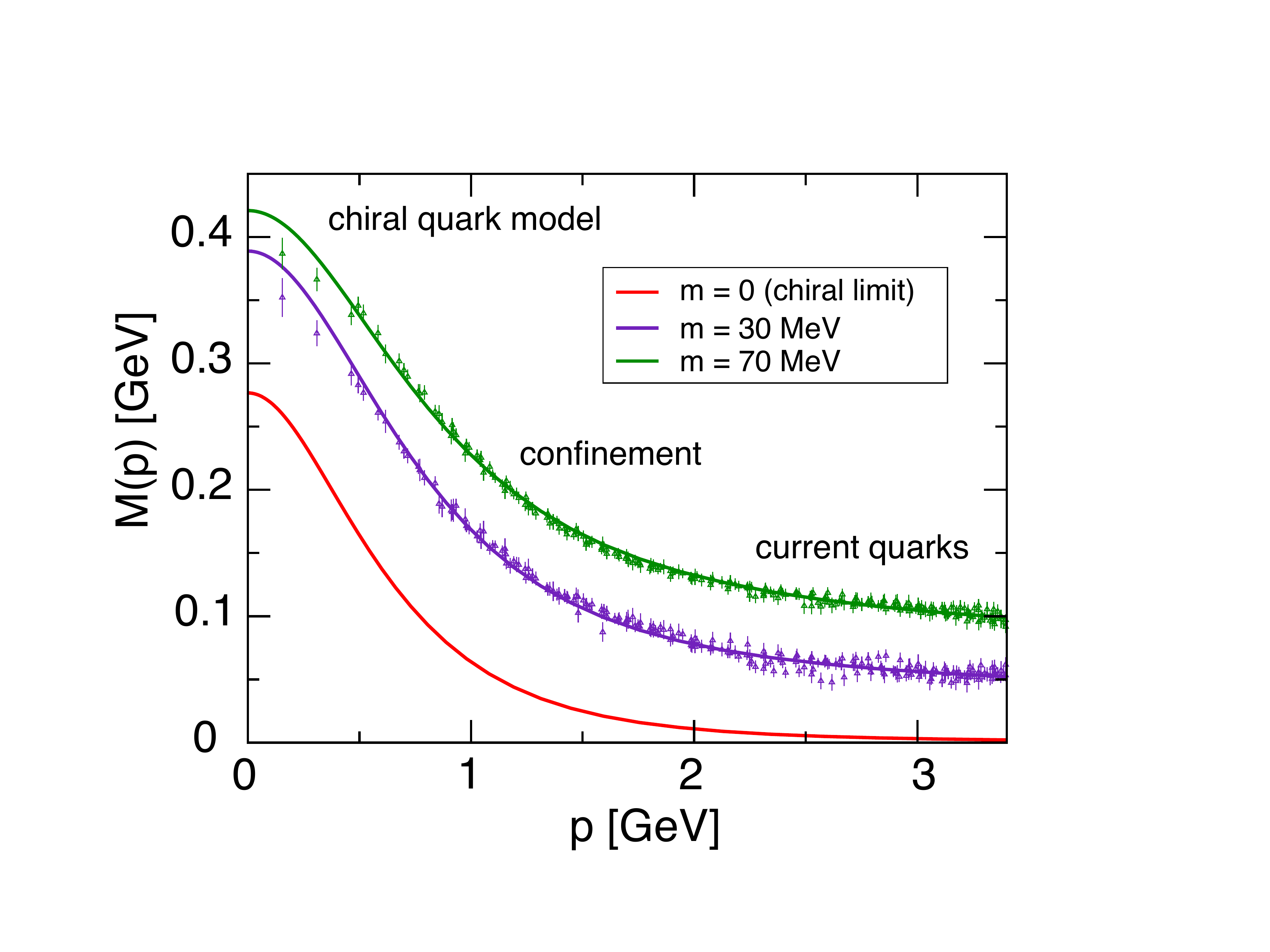}
    \caption{The dressed-quark mass function $M(p)$,  computed by Dyson--Schwinger techniques, compared with numerical simulations of lattice-QCD for several values of the current-quark mass $m$ (adapted from \cite{Bhagwat:2007vx}).}
    \label{fig:uni}
  \end{center}
\end{figure}

The chiral quark model presents an intuitive picture of how quasistatic properties of the nucleon might arise. Perturbative evolution of the parton distributions treats symmetrically quarks and antiquarks, and up and down flavors. Any differences between quark and antiquark distributions, or the distributions of up and down quarks, must therefore be set at low scales. An interesting challenge is presented by the Gottfried sum rule,
\begin{eqnarray}
I_{\mathrm{G}}(Q^2) & \equiv & \int_0^1 dx \frac{F_2^{p}(x,Q^2) - F_2^{n}(x,Q^2)}{x} \nonumber \\
 & = & \int_0^1 dx \sum_i e_i^2\left[q_i^{(p)}(x,Q^2) + \bar{q}_i^{(p)}(x,Q^2)
 q_i^{(n)}(x,Q^2) + \bar{q}_i^{(n)}(x,Q^2)\right],
\label{eq:gottsr}
\end{eqnarray}
for which the New Muon Collaboration long ago found~\cite{Amaudruz:1991at} the value $0.240 \pm 0.016$, below the expectation of $\frac{1}{3}$ in the quark-parton model with a flavor-symmetric light-quark sea. To see how an asymmetry could come about, imagine that a constituent quark fluctuates into a current quark plus a Nambu--Goldstone boson, as illustrated in Figure~\ref{fig:chiqm}~\cite{Eichten:1991mt}.
\begin{figure}
\begin{center}
{\includegraphics[width=0.9\textwidth]{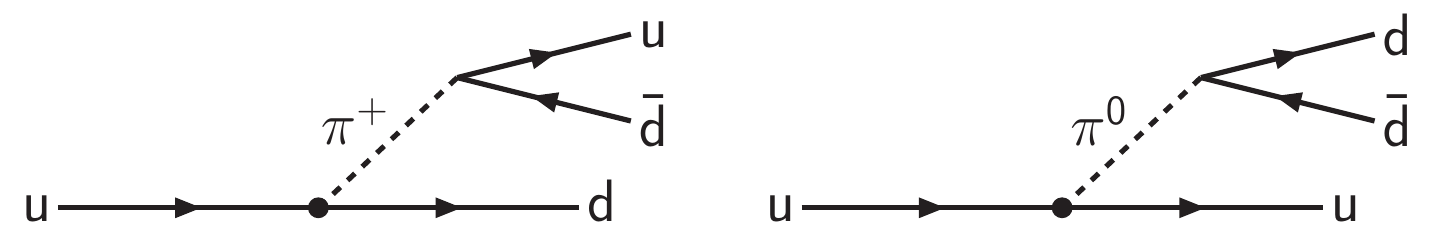}}
\caption{Example fluctuations of a valence up-quark in the chiral quark model. }
\label{fig:chiqm}
\end{center}
\end{figure}
The resulting pion cloud changes the parton distributions for protons and neutrons, but doesn't enter the difference $F_2^{p} - F_2^{n}$ because $F_2^{\pi^+} = F_2^{\pi^-}$. The Gottfried sum rule defect arises from the asymmetry of the left-behind quarks, which arises from the $(uud)$ vs. $(udd)$ composition of valence quarks in the proton and neutron. The pion cloud also doesn't affect the spin budget of the nucleon, because the pion is spinless. However, the $\gamma_5$ coupling flips the helicity of the left-behind quark. Consequently, we should expect spin contributions $\Delta d, \Delta s < 0$, $\Delta{\bar{d}}, \Delta{\bar{s}} = 0$, which set interesting targets for experiment.

Even at low scales, where the strong interactions are \textit{a priori} strong, we know cases in which partonic arguments yield results in agreement with experimental observations. One example is the surprising success of the simple picture of hadron masses put forward by De Rujula, Georgi, and Glashow~\cite{DeRujula:1975ge}.
An analytic hadronization model for event shapes by Dokshitzer, Marchesini and Webber  derives the structure of power corrections from perturbative QCD~\cite{Dokshitzer:1995qm}. The model assumes that the strong coupling remains finite at low energy scales where simple perturbative calculations break down. A mean value at low scales,
\begin{equation}
\alpha_0(\mu_I) \equiv (1/\mu_I)\int_0^{\mu_I}dQ \,\alpha_{\mathsf{s}}(Q),\quad \mu_I = 2\gev
\label{ep:alfnought}
\end{equation}
has been extracted from several observables~\cite{MovillaFernandez:2001ed}. These are consistent with the idea that the effective coupling freezes at low scales, with $\alpha_0(2\gev) \approx 0.5$. If that is indeed the case, it is plausible to consider quarks and gluons as apt long-distance degrees of freedom. It may also clarify such noteworthy simplicities as the unimportance of nonvalence components for hadron properties, Bloom--Gilman duality, precocious scaling, and the perturbative approach to bound states~\cite{Hoyer:2011nz}.

\subsection{Could chiral symmetry and confinement coexist?}
Confinement and the spontaneous breaking of chiral symmetry for light quarks go hand in hand~\cite{Cheng:2007jq}. In the case of heavy--light $(Q\bar{q})$ mesons and doubly heavy $(QQq)$ baryons, it is interesting to inquire whether a vestige of the chiral symmetry of the tethered light quark might be revealed in parity-doubled bound states transforming as linear representations of the light-quark chiral symmetry~\cite{Bardeen:2003kt}. For the heavy--light mesons $D$, $D_s$, $B$, $B_s$, we should then expect chiral supermultiplets linking states with adjacent values of the orbital angular momentum, $(L, L+1)$, but with the same value of the light-quark spin, $j_q$, that we encountered in \S\ref{sec:control}. In particular, the chiral symmetry should link the levels
\begin{eqnarray*}
j_q = {\textstyle\frac{1}{2}}: \quad \mathrm{1S}(0^{-},1^{-})\hbox{ and }\mathrm{1P}(0^{+},1^{+}) \\
j_q = {\textstyle\frac{3}{2}}: \quad \mathrm{1P}(1^{+},2^{+})\hbox{ and }\mathrm{1D}(1^{-},2^{-}) \\
\ddots\phantom{MMMMMMM}
\end{eqnarray*}
This association of parity doublets implies that the hyperfine splitting should be the same for partner levels: $M_{1^+}-M_{0^+} = M_{1^-} - M_{0^-}$, etc. It will be very interesting to extend observations to the doubly heavy baryons and to establish for a range hadrons with tethered light quarks how far QCD is from the conjecture described here.

\subsection{Proton structure and dark matter searches}
The search for cold dark matter as weakly interacting massive particles (WIMPs) is advancing on many fronts~\cite{taup}. Passive experiments that seek direct detection  by registering collisions of passing WIMPs with the detector material provide limits or signals expressed in terms of the WIMP mass and the cross section for spin-dependent or spin-independent scattering. There is high interest in confronting observations with the expectations of hypotheses about the nature of the dark-matter particles~\cite{Feng:2010ef}. That is where hadronic physics enters. For popular forms of neutralino dark matter, the dominant contribution to the spin-independent cross section plausibly arises from Higgs-boson exchange between the WIMP and the target nucleon. How does the Higgs boson interact with the nucleon? Surely not by coupling to the nucleon mass, because QCD has taught us that the nucleon mass is generated to very high approximation by the strong interactions. Higgs couplings to the up and down quarks are small, in proportion to their tiny current-quark masses, so it is likely that the major contribution arises from Higgs couplings to the minority ``hidden'' constituents of the nucleon, strange quarks and beyond. The role of $(\bar{s}s)$~\cite{Babich:2010at} and heavier condensates and details of the gluon condensate are incompletely understood, and this gives rise to an uncertainty in the spin-independent cross section of up to an order of magnitude~\cite{Giedt:2009mr}. We can certainly hope to address the narrow question of heavy-flavor presence in the nucleon both experimentally and theoretically. Perhaps now is the time to ask whether a theoretical reconception is called for. 

\subsection{Nucleon structure at a neutrino factory?}
If a millimole of muons produced in hadron reactions can be captured, cooled, and accelerated to some tens of GeV, it may be possible to construct a neutrino factory, for which the instability of the stored muons is a virtue~\cite{vlenf}. The intense neutrino beams would be ideal for the study of neutrino oscillations over long baselines, but the possibilities of close-in experimentation are also quite remarkable. With a flux of $10^{20}$ neutrinos per year, we could contemplate $\nu$ scattering on thin targets, \eg, hydrogen or deuterium, $\nu$ scattering on active semiconductor targets to study heavy-flavor production, and even $\nu$ scattering on polarized targets. Preliminary studies indicate that it would be possible to determine flavor by flavor the valence and sea-quark distribution functions with very small statistical errors~\cite{Ball:2000qd} and to extract detailed information on polarized structure functions~\cite{Forte:2005pv}. A modern critical evaluation of these conclusions would be very enlightening.

\section{Spectroscopy of states containing heavy quarks}
We have extensive knowledge of baryons containing a single $c$ or $b$ quark, but no established observations of doubly heavy baryons. Searching for $QQq$ states is an important goal for the LHC experiments. Let us review the essential expectations. In the heavy-quark limit, there are parallels between $[QQ^{(\prime)}]_{\mathbf{3^*}}q$ baryons and $Q\bar{q}$ mesons as heavy--light systems. At the level of one-gluon-exchange, 
\begin{equation}
V_{[QQ^{(\prime)}]_{\mathbf{3^*}}}(r) = {\textstyle{\frac{1}{2}}}V_{(Q\bar{q})_{\mathbf{1}}}(r).
\end{equation}
How faithfully does this simple estimate reflect reality? In addition to the meson-like excitations of the tethered light quark, the doubly heavy baryons should exhibit excitations of the core, from which we may learn about $[QQ^{(\prime)}]_{\mathbf{3^*}}$ dynamics. If we are able to explore $bcq$ baryons, the unequal masses of the heavy quarks might give us a window on shortcomings of intuition drawn from nonrelativistic potential models, just as we hope for the $B_c$ mesons.

Weak decays of doubly heavy baryons entail a rich set of heavy-heavy and heavy-light transitions that isolate different pieces of the effective Hamiltonian for charged-current interactions. Strong and electromagnetic cascades from excited states will involve two length scales, $r_{\mathrm{H}} = \langle r^2_{(QQ^{(\prime)})}
 \rangle^{\half}$ and $r_\ell = \langle r^2_{(QQ^{(\prime)}q)}
 \rangle^{\half}$. The core excitations in particular should reveal some extremely narrow states. Studies of production dynamics should also stretch our understanding, both by extending fragmentation models to new regimes and through the comparison with quarkonium production dynamics.

For the $B_c$ meson family, I will be very brief. We need to elaborate the experimental systematics of decays, including the $\tau \nu_\tau$ annihilation channel, $b$ decays, and $c$ decays. It would also be highly informative to reconstruct part of the excitation spectrum in $\pi^+\pi^-$ or $\gamma$ cascades.

Finally, a few words on the ordinary quarkonium states of the $\psi$ and $\Upsilon$ families and the new states associated with quarkonium. For the ordinary states, we should aim for a comprehensive understanding of the decays to $\pi^+\pi^-(Q\bar{Q})$. There are enough inconsistencies and puzzling effects that it is time to rethink our reliance on the color multipole expansion. This has implications for some of the extraordinary states as well---those that have mysteriously large branching fractions to $\pi^+\pi^-(Q\bar{Q})$.

The many new states associated with charmonium raise many new questions and opportunities. A few [$\chi_{c2}(2P)(3927)$, for example] look like simple $c\bar{c}$ levels. Most new states are not simple quarkonium, and indeed we should expect most states above the flavor threshold to have multiple personalities. Not all the expected charmonium levels have yet been observed. In addition to the $\eta_{c2}(1\slj{1}{3}{2})$ and $\psi_2(1\slj{3}{3}{2})$, the narrow states 1\slj{3}{3}{3}, 2\slj{3}{2}{2}, and 1\slj{3}{4}{4} remain at large. The rich harvest of states, including candidate charged states that do not fit the standard $Q\bar{Q}$ body plan, shows the virtue of making all possible few-particle combinations.

On the theory side, we need to better understand the role of thresholds in shaping the states associated with quarkonium on their own, near would-be quarkonium levels, and in the company of attractive $s$-wave amplitudes. Close scrutiny of the similarities and differences of the new states associated with $(c\bar{c})$ and $(b\bar{b})$ should be very informative.

\section{A New Era of Heavy-Ion Physics}
The 2010 heavy-ion run of the LHC, devoted to Pb-Pb collisions at $\sqrt{s_{NN}} = 2.76\tev$, has begun to realize the great potential of experimentation in the new energy regime to illuminate a broad array of topics in hadronic physics. Effects long sought, or teased out of the data with great difficulty at lower energies, show themselves readily at the LHC. The new challenge is to develop the new observations into robust and quantitative tools. I give two illustrations. 

If a hard-scattering event occurs within a  quark-gluon plasma, the outgoing partons will lose energy through collisions with the hot stuff of their environment~\cite{Bjorken:1982tu}. Should one of the partons traverse a longer path in the hot matter, it would suffer a greater energy loss than its partner; the ensuing jet of hadrons would be degraded, or even extinguished. Both ATLAS~\cite{Aad:2010bu} and CMS~\cite{Chatrchyan:2011sx} have observed, and begun to characterize, just such an effect.

A quark-gluon plasma is expected to screen the potential that binds heavy quark-antiquark pairs and so to melt quarkonium states---the lightly bound excited states first~\cite{Matsui:1986dk}. The CMS Collaboration reports the double ratio~\cite{Chatrchyan:2011pe}
\begin{equation}
\frac{\left.{\Upsilon(\mathrm{2S+3S})}/{\Upsilon(\mathrm{1S})}\right|_{\mathrm{[Pb\,Pb]}}}{\left.{\Upsilon(\mathrm{2S+3S})}/{\Upsilon(\mathrm{1S})}\right|_{[pp]}} = 0.31_{-0.15}^{+ 0.19}\hbox{ (stat.)} \pm 0.03\hbox{ (syst.)}.
\label{eq:fondre}
\end{equation}
Analysis of the sequential-melting scenario is complicated by the fact that a significant fraction of the observed yield of the deeply bound $\jpsi(\mathrm{1S})$ and $\Upsilon(\mathrm{1S})$ states is due to decays of excited states. It would therefore be of great interest to investigate directly the melting of $\chi$ states, and to compare the behavior of the $\jpsi$ and $\Upsilon$ families. The energy dependence of the phenomenon should provide insight into the nature of the hot medium. Studies of quarkonium melting in the $B_c$ system should also be enlightening, once excited states have been observed.

We can expect much more from ALICE, ATLAS, and CMS in the next heavy-ion run.

\section{Concluding Remarks}
    
QCD has been validated as a new law of nature. It is internally consistent up to very high energies, and so could be a complete theory of the strong interactions. Whether QCD is the final answer for the strong interactions is a subject for continuing experimental tests, which are  being extended in  experimentation at the Large Hadron Collider. Beyond the comparison of perturbative calculations with experiment, it remains critically important to test the confinement hypothesis by searching for free quarks, or for signatures of unconfined color. Sensitive negative searches for quarks continue to be interesting, and the definitive observation of free quarks would be revolutionary. Breakdowns of factorization would compromise the utility of perturbative QCD. Other discoveries that would require small or large revisions to QCD include the observation of new kinds of colored matter beyond quarks and gluons, the discovery that quarks are composite, or evidence that \cgg\ gauge symmetry is the vestige of a larger,
spontaneously broken, color symmetry. 

While probing our underlying theory for weakness or new openings, we have plenty to do to apply QCD to myriad experimental settings, to learn its implications for matter under unusual conditions, and to become more adept at calculating its consequences. New experimental tools provide the means for progress on a very broad front.

\acknowledgements{%
It is a pleasure to express heartfelt thanks to the International Advisory Committee and the Local Organizing Committee, and especially to Stephan Paul and Karin Frank, for preparing a superb program and a stimulating week in the inspiring surroundings of the K\"{u}nstlerhaus. I am grateful to the Excellence Cluster \textit{\guillemotleft Origin and Structure of the Universe\guillemotright} for generous support. I thank Eric Laenen and the NIKHEF Theory Group for warm hospitality during the preparation of this manuscript. I salute the contributors and participants for excellent presentations and enlightening discussions. Fermilab is operated by the Fermi Research Alliance under contract no.\  DE-AC02-07CH11359 with the U.S.\ Department of Energy.

}


%

}

\end{document}